\documentclass[preprint,aps,12pt,showpacs,showkeys,nofootinbib]{revtex4}

\usepackage{graphicx}
\usepackage{epstopdf}
\usepackage{dcolumn}
\usepackage{amsfonts}
\usepackage{amsmath}
\usepackage{amssymb}
\usepackage{bm}
\usepackage{color}
\usepackage{comment}
\newcommand{\dsfrac}{\displaystyle\frac}

\begin{document}	
	
	\title{Characteristic temperatures of a triplon system of dimerized quantum magnets}
	\author{Abdulla Rakhimov$^{a}$}\email{rakhimovabd@yandex.ru}
	\author{Mukhtorali Nishonov $^{b}$}\email{nishonov@inp.uz}
	\author{Luxmi Rani$^{a}$}\email{luxmi.rani@bilkent.edu.tr}
	\author{B. Tanatar $^{a}$}\email{tanatar@fen.bilkent.edu.tr}	
	
	\affiliation{$^a$Department of Physics, Bilkent University, Bilkent, 06800 Ankara, Turkey. \\
		$^b$Institute of Nuclear Physics, Tashkent 100214, Uzbekistan}

\begin{abstract}
Exploiting the analogy between ultracold atomic gases and the system of triplons, we study magneto-thermodynamic properties of dimerized quantum magnets in the framework of  Bose -Einstein condensation (BEC). Particularly, introducing the inversion (or Joule - Thomson) temperature $T_{JT}$ as the point where Joule - Thomson coefficient
of an isenthalpic process changes its sign, we show that for a simple paramagnet, this temperature is infinite, while for three-dimensional (3D) dimerized quantum magnets it  is  finite and always larger than the critical temperature $T_c$ of BEC. Below the inversion temperature  $T<T_{JT}$  the system of triplons may be in a liquid phase, which undergoes a transition into a superfluid phase at
$T\le T_c<T_{JT}$. The dependence of the inversion temperature on the external magnetic field $T_{JT} (H)$ has been calculated for quantum magnets of TlCuCl$_3$  and  Sr$_3$Cr$_2$O$_8$.
\end{abstract}
   \maketitle

\section{Introduction}
The properties of dimer spin systems at low temperatures have been intensively investigated in the last two decades. These magnetic systems, e.g., TlCuCl$_3$, Sr$_3$Cr$_2$O$_8$, etc.\cite{zapf} consist of weakly coupled dimers with strong antiferromagnetic interaction between spins within a dimer. The ground state in such components is singlet and it is separated from the first exited triplet state by a
gap at zero magnetic field at zero temperature that may be interpreted as a spin-liquid behavior characterized by a finite correlation length \cite{ruegg}.
When an external magnetic field $H$ is applied, the gap can be closed due to the Zeeman effect, resulting in the generation of a macroscopic number of triplet excitations (triplons) and the transition to a magnetically ordered phase takes place at $H=H_c$. This transition has been observed by studying the magnetization  of e.g., TlCuCl$_3$ nearly 20 years ago \cite{tanaka}. Further, it was shown that it may be effectively described in terms of Bose-Einstein condensation (BEC) of quasi-particles
of triplons \cite{yamada,cavadini},  which mathematically  can be introduced by a generalized Schwinger representation in the bond-operator formalism \cite{sachdev,yukalovtriplon}. In a constant external magnetic field and zero temperature the number of triplons is conserved in the thermodynamic limit and controlled by  an effective \cite{yukalovtriplon,radu,mills}   chemical potential $\mu$ defined as
\begin{eqnarray}
\mu&=&g_f\mu_B(H-H_c),\label{mu}
\end{eqnarray}
where $g_f$ is electron Lande factor and $\mu_B$ is the Bohr magneton.

A triplon does not carry mass or electric charge, but a magnetic moment. So, it can be easily understood that the total number of triplons, $N$ defines the uniform magnetization $M$, while the number of condensed triplons $N_0$ defines the staggered magnetization $M_{stag}$, namely \cite{tanaka}
\begin{eqnarray}
M&=&g_f\mu_B N,
\label{maguni}\\
M_{stag}&=&g_f\mu_B\sqrt{\frac{N_0}{2}}.
\label{magstag}
\end{eqnarray}
Here it should be noted that, in the thermodynamic limit, BEC is accompanied by spontaneous breaking of global gauge symmetry, which is  a necessary and sufficient condition \cite{yukalovtriplon}. But in real materials, e.g. in TlCuCl$_3$, this symmetry can be explicitly broken due to anisotropy. As a result, instead of a phase transition one has to deal with a crossover where the staggered magnetization is renormalized \cite{sirker,ouraniz1,kolezhuk,ourann1,ourann2}. In the present work, for simplicity, we shall neglect such effects and exploit Eqs.\,(\ref{maguni})  and (\ref{magstag}).

The investigation of analogy between ordinary gases and the system of magnons has been made  by Bovo et al. \cite{bovo}. Studying frustrated ferromagnets, they have found that, analogous to gases, magnets have at least two kinds of critical temperatures, namely Joule $T_J$ and Joule-Thomson $T_{JT}$ temperatures. By definition $T_J$ corresponds to the temperature for which the system is quasi-ideal and the internal energy $E$ is independent of the the extensive parameters like volume (c.f. Table I of Ref.\cite{bovo})
$\left({\partial E}/{\partial V}\right)_T=0$, or magnetization
$\left({\partial E}/{\partial M}\right)_T=0$. As to the $T_{JT}$, it is related to the well known Joule-Thomson isenthalpic process which is characterized by the following  coefficient
\begin{eqnarray}
\kappa_{JT}&=&\left\{
\begin{array}{l}
\left(\frac{\partial T}{\partial P}\right)_W=\frac{1}{C_P}
\left[T\left(\frac{\partial V}{\partial T}\right)_P-V\right],\,\,\text{gases}\\
\\
\left(\frac{\partial T}{\partial H}\right)_W=\frac{1}{C_H}\left[M-T\left(\frac{\partial M}{\partial T}\right)_H\right]\,\,  \text{, paramagnets}.
\end{array}
\right.
\label{kappajt}
\end{eqnarray}
where $C_P$ and $C_H$ are heat capacities at constant pressure and magnetic field , respectively. The sign of $\kappa_{JT}$ indicates whether the system heats up ($\kappa_{JT}>0$) or cools ($\kappa_{JT}<0$) during the process when the intensive parameter, $P$ or $H$ is increased. By definition the inversion temperature is the temperature when $\kappa_{JT}$ changes its sign i.e., $\kappa_{JT}(T=T_{JT})=0$. Note that for a classical ideal gas $\kappa_{JT}=0$ at any temperature whereas ideal quantum gases have non-zero $\kappa_{JT}$ at low temperature \cite{sisman}.  Such quantum isenthalpic process has been recently observed  in a saturated homogeneous Bose gas \cite{Schmidutz}.

In practice $T_{JT}$ shows the starting of the regime below which a gas may be liquefied by the Linde-Hampson isenthalpic process. For example for helium $T_{JT}=34$\,K, which means that one has to cool helium until 34\,K to obtain liquid helium using the Joule-Thomson effect. In Refs. \cite{aczelprl103,wangprl116} it has been argued that, a three-dimensional (3D) spin-dimerized quantum magnet exhibits a triplon-superfluid phase between $H_{c_1}$ and $H_{c_2}$ (saturation field). This superfluid phase is embedded in a dome-like phase diagram of triplon liquid extending up to $T_c^{max}$, maximum temperature of the magnetically ordered regime \cite{wangprl116,barmaley}, as it is illustrated in Fig.\,4 of Ref.\,\cite{wangprl116}. Particularly,  $T_c^{max}\leq 9$\,K  both for
Sr$_3$Cr$_2$O$_8$ and TlCuCl$_3$.

As discussed by Wang et al. \cite{wangprl116} the ground-state of such a system is a quantum disordered paramagnet with spin gapped elementary excitation, triplon. When Zeeman energy compensates the intra-dimer interaction, a QPT from quantum disordered (QD) phase to a spin aligned state can be induced. The paramagnetic and ferromagnetic (FM) states are separated by a canted-XY antiferromagnetic (AFM) phase, which can be viewed as a triplon superfluid. The superfluid fraction survives up to $T_c^{max}\approx 8$\,K  and the triplon exhibit liquid-like behavior up to
$T^*\sim 18$\,K, as it was confirmed by analyzing the sound velocity measurements. Now, coming back to the analogy with ordinary atomic systems, we may propose that in spin-dimerized magnets $T_c$ corresponds to the critical temperature of BEC, while $T_{JT}$ corresponds to $T^*$  of Ref.\,\cite{wangprl116}, i.e., to the temperature below which triplons may be considered as a liquid. In other words, we assume that similarly to ordinary gases, $T_{JT}$ is the temperature, when for $T>T_{JT}$ triplon gas can not be "liquefied".

Therefore, the main purpose of the present work is to estimate magnetic analogies of such critical temperatures, $T_c$, $T_J$, and $T_{JT}$  in spin gapped  magnets.
\footnote{Namely, $T_c$ - critical temperature of BEC; $T_J$ - Joule temperature when the gas behaves as an ideal gas; $T_{JT}$ - inversion temperature, such that $\kappa_{JT}(T)=0$; $T^{*}$ is the maximal temperature, below which magnons can be considered in a liquid phase, as predicted in \cite{wangprl116}}

The rest of the paper is organized as follows. In Sect. II we present general analytical expressions of magnetic thermodynamics.  In Sect.  III we discuss our predictions concerning $T_J$ and $T_{JT}$. The main conclusions are drawn in Sect. IV. The details of some calculations are presented in the Appendices {\bf  A} and {\bf B}.

\section{Basic relations of magnetic thermodynamics}

Generally speaking, the total Hamiltonian (or energy) of a magnetic substance is usually assumed to consist of several contributions: from the crystalline lattice ($\hat{H}_L$) and from the conducting electrons ($\hat{H}_e$), besides, the magnetic moments ($\hat{H}_m$) and from the atomic nucleus ($\hat{H}_n$). So are thermodynamic potentials, e.g. the grand potential $\Omega$ and the entropy, $S$. For the sake of simplicity, we assume that $\Omega_L$ and $\Omega_e$ do not depend on the applied magnetic field and,  hence the total changes induced by the magnetic field variation are attributed to the changes of only the magnetic part. Below we concentrate only on the magnetic part of a physical variable denoting e.g., $\Omega_M$ as just $\Omega$: $\Omega\equiv \Omega_M$. In the next section we derive $\Omega$ explicitly for spin gapped magnets while here we present some general relations, assuming that $\Omega$  is known.

Thus, we have the following relations for main thermodynamic potentials \cite{landau}
\begin{eqnarray}
\begin{array}{l}
F=\Omega+\mu N,E=F+TS, \quad \Phi=W-TS=\mu N\\
\\
W=E+PV-HM=\mu N+TS,
\end{array}\label{eq52}
\end{eqnarray}
where $E$, $\Phi$, $W$ and $F$ are internal energy, Gibbs free energy, enthalpy and Helmholtz potential, respectively. The total differentials are \cite{yukalovtutor,nolting}
\begin{equation}
\begin{array}{lcl}
d\Omega&=&-SdT-PdV-Nd\mu-MdH,\\
dF&=&-SdT-PdV+\mu dN+HdM,\\
dE&=&TdS-PdV+\mu dN+HdM,\\
d\Phi&=&-SdT+VdP+\mu dN-MdH,\\
dW&=&TdS+VdP+\mu dN-MdH\ .
\end{array}
\label{eq61}
\end{equation}

Now, passing to the discussion of temperatures $T_J$ and $T_{JT}$, it can be shown (see Appendix A) that $T_J$ corresponds to a local extremum of the quantity $\chi T$, i.e.,
\begin{eqnarray}
\frac{d}{dT}(\chi T)\Big|_{T=T_J}&=&0,\label{eq62}
\end{eqnarray}
where we defined the susceptibility as\footnote{The Eq. (\ref{sus}) should be considered just as a notation, not a linear approximation, which holds for a weak magnetic field.}
\begin{equation}
\chi\equiv \frac{M}{H}.
\label{sus}
\end{equation}
which still depends on the magnetic field , $\chi=\chi(H)$  .
Equation\,(\ref{eq62}) may be  represented in following equivalent form
\begin{equation}
\left[M +T\left(\frac{\partial M}{\partial T}\right)_H\right]\Big|_{T=T_j}=0\, .
\label{eq666}
\end{equation}
Therefore, studying the temperature dependence of a physical observable such as the magnetic susceptibility $\chi(T,H)$ one may pinpoint the Joule temperature, $T_J$, where the triplon (or magnon) system behaves like a quasi-ideal system.

An isenthalpic process ($W={\rm const}.$) being a main part of Joule-Thomson effect is characterized by the Joule-Thomson coefficient $\kappa_{JT}\equiv\left({\partial T}/{\partial H}\right)_W$ (similar to $\kappa_{JT}\equiv\left({\partial T}/{\partial P}\right)_W$ for atomic gases). As it was shown in Appendix A $\kappa_{JT}$ can be represented as
\begin{eqnarray}
\kappa_{JT}&=&
\frac{1}{C_H}\left[M-T\left(\frac{\partial M}{\partial T}\right)_H\right]\, .\label{eq72}
\end{eqnarray}

Finally, the inversion temperature $T_{JT}$ is the solution of $\kappa_{JT}(T=T_{JT})=0$, which leads to
\begin{eqnarray}
\frac{d(\chi/T)}{dT}\Big|_{T=T_{JT}}&=&0.\label{eq83}
\end{eqnarray}
Using Eqs.\,(\ref{sus}) and (\ref{eq72}) we can see that at the inversion temperature $T_{JT}$ the quantity $\chi/T$ has a local extremum, i.e., ${d(\chi/T)}/{dT}$ changes its sign. Equations\,(\ref{eq52})-(\ref{eq83}) are general for any paramagnetic material. In the next section we derive thermodynamic quantities specifically for spin gapped dimerized quantum magnets.

\section{Results and discussions}

For simplicity we start with a paramagnetic material whose magnetization is given as \cite{kittel}
\begin{eqnarray}
M&=&g_f\mu_B\tanh(x)\label{eq11p1}
\end{eqnarray}
where $x=g_f\mu_B H/T$. From equations (\ref{eq72}) and (\ref{eq11p1}) one easily obtains
\begin{equation}
\kappa_{JT}(x)=\frac{g_f\mu_B}{C_H}
\left[\tanh(x)+\frac{x}{\cosh^2(x)}\right]
\label{eq11p3}
\end{equation}
It is clear that in this case $\kappa_{JT}(x)=0$ corresponds to $x=0$, that is
the inversion tempearture  $T_{JT}$ (paramagnetic)$\to$ $\infty$, which means
that a magnon fluid in paramagnetic materials can never be considered as a liquid at finite temperature. As to the Joule temperature defined in Eq.\,(\ref{eq62})
it is easy to show that $T_J$ is also infinite, which can be proven using the Eqs.\,(\ref{eq62}), (\ref{sus}) and (\ref{eq11p1}).

Now passing to dimerized quantum magnets,  we adopt commonly used set of realistic parameters $g_f$, $H_c$, $U$ and $J_0$, which have been  fitted to experimental data for Sr$_3$Cr$_2$O$_8$ and TlCuCl$_3$ \cite{aczelprl103,delamore,kohama}, as presented in Table I.

\begin{center}
	\begin{table}
		
		{\begin{tabular}{|p{4cm}|c|c|c|c|c|c|}
			\hline
			   &$g_f$& $H_c$ (T) &$J_0$ (K) &$U$ (K)\\
			\hline
			Sr$_3$Cr$_2$O$_8$&1.95&30.4&15.86&51.2\\
			\hline
			TlCuCl$_3$&2.06&5.1&50&315\\
			\hline
		\end{tabular}}
	    \caption{Material parameters used for our numerical calculations.
	    	From the experimental input parameters $g_f$ and  $H_{c}$,  we derived $J_{0}$
	    	and  coupling constant $U$  by fitting the experimental phase boundary $T_{c}(H)$ to Eqs.(\ref{mu}) and  (\ref{maguni}) (see Ref. \protect\cite{ourmce}  for the details). Note that, here  all quantities are given in the units with
	    	$k_B=\hbar=V=1$.}
		\label{T:result 2}
	\end{table}
\end{center}
These parameters are included in following effective Hamiltonian
\begin{eqnarray}
H&=&\int d\vec{r}\left\{\Psi^\dagger\left[\hat{K}-\mu\right]\Psi+
\frac{U}{2}(\Psi^\dagger\Psi)^2\right\}
\label{ham}
\end{eqnarray}
where $\Psi$ is the bosonic field, $\mu$ is the chemical potential given in
Eq.\,(\ref{mu}), and $U$ is a coupling constant of triplon-triplon contact interaction, which is usually considered as a fitting parameter. The kinetic energy operator, $\hat{K}$ gives rise to the bare disperison
$\varepsilon_k=J_0(3-\cos ak_x-\cos ak_y-\cos ak_z)$.

Now we  discuss the inversion temperature $T_{JT}$ of these compounds \footnote{Details of calculation of magnetizations , heat capacity etc  could be found in our work \cite{ourmce}}. In Figs. 1(a,b) we present Joule - Thomson coefficient for Sr$_3$Cr$_2$O$_8$ (a) and TlCuCl$_3$ (b). As it is seen from Figs.\,1(a,b) magnetic Joule-Thomson coefficient  $\kappa_{JT}$ crosses the  abscissa at a moderate value of the temperature. Therefore, in contrast to a simple paramagnet,
the inversion temperature for dimerized magnets is finite. To study this point in more detail we shall  look for  a  possible extremum of  the function  $\chi(T,H)/T$, in accordance with the Eq.\,(\ref{eq83}). In Figs. 2(a,b) we present
${d (\chi/T)}/{dT}$ vs temperature for Sr$_3$Cr$_2$O$_8$  $(H=33\,T)$  and TlCuCl$_3$, $(H=6\,T)$, respectively. It is seen that ${d (\chi/T)}/{dT}$ changes its sign at temperatures higher than critical one, $T_{JT}>T_c$.

We address the question of information that can be extracted from experiments, say, from the extremum of the function $\chi/T$,  which is related to $M(T,H)$.
Unfortunately, there is no experimental data on $M(T)$ available for $Sr_3Cr_2O_8$, but there is a plenty of data on $M(T)$ for $TlCuCl_3$ \cite{yamada,delamore}.
So,  we adopted the existing data on $M(T,H)$ for this material, e.g., given in
Ref.\,\cite{delamore} and using Eq.\,(\ref{sus}), we constructed the dependence of
${d (\chi/T)}/{dT}$ on temperature. From Fig.\,2b we see that the experimental value of $T_{JT}$ for TlCuCl$_3$ at $H=6T$ is $T_{JT}^{exp} (H=6T)\approx 3.9K$. This fact  confirms  the existence of a finite inversion temperature for the compound TlCuCl$_3$, which has no frustration. As to our theoretical prediction, it is seen that, the solid line in Fig. 2(b) crosses the abscissa at  a larger temperature, approximately at  $T_{JT}^{HFB} (H=6T)\approx 5K$. It appears that our estimate is in good qualitative agreement with the experiment.
\begin{figure}[h!]
	\begin{minipage}[H]{0.49\linewidth}
		\center{\includegraphics[width=1.0\linewidth]{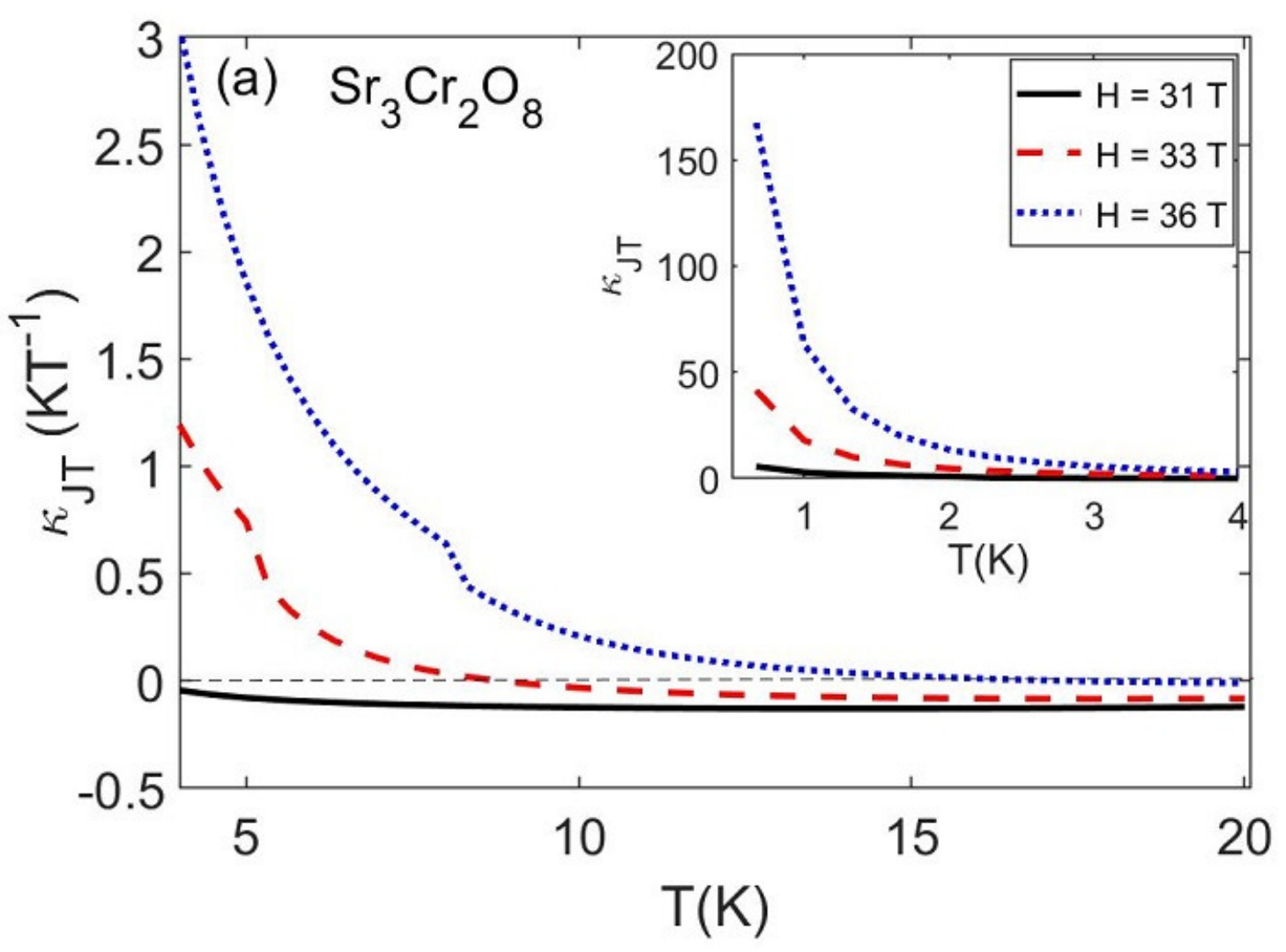} \\ a)}
	\end{minipage}
	\hfill
	\begin{minipage}[H]{0.49\linewidth}
		\center{\includegraphics[width=1.0\linewidth]{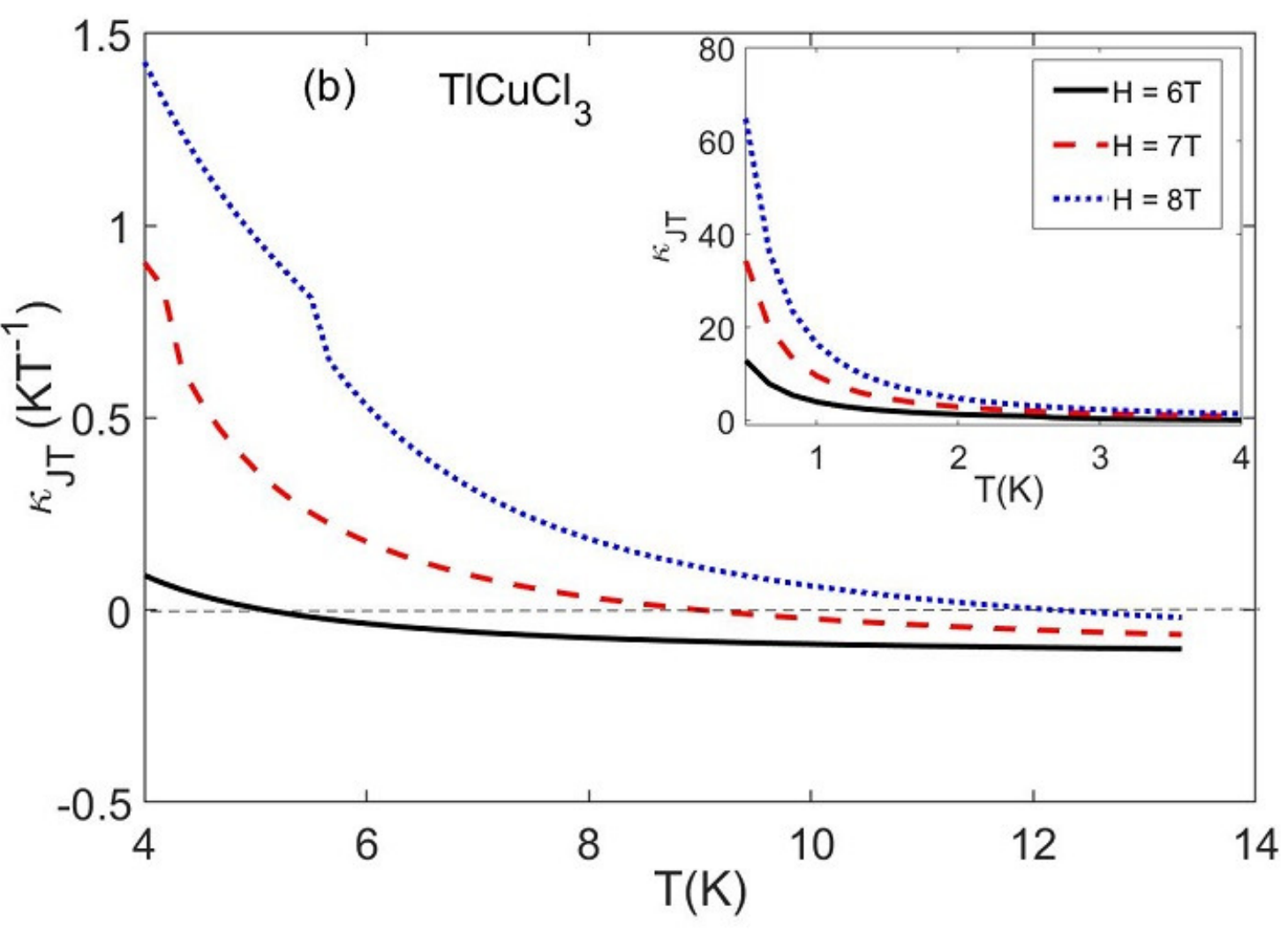} \\ b)}
	\end{minipage}
	\caption
	{
		The temperature dependence of the Joule - Thomson coefficient for Sr$_3$Cr$_2$O$_8$ (a) and
		TlCuCl$_3$ (b). The point where $\kappa_{JT}$  crosses absicca correspond to inversion temperature
		for each magnetic field.
		Inset: $\kappa_{JT}$  for small values of T.
	}
	\label{fig1}
\end{figure}

Similarly to the inversion temperature of atomic gases, which depends on pressure,
the inversion temperature of a magnetic Joule - Tomson process depends on the external magnetic field, which is presented in Figs\,3(a,b). As it is seen, for both materials this temperature is larger than the critical temperature of BEC, and
the dependence of the dimensionless ratio $T_{JT}/T_c$ on the magnetic field is rather small.

\begin{figure}[htbp!!]
	\begin{minipage}[H]{0.49\linewidth}
		\center{\includegraphics[width=1.0\linewidth]{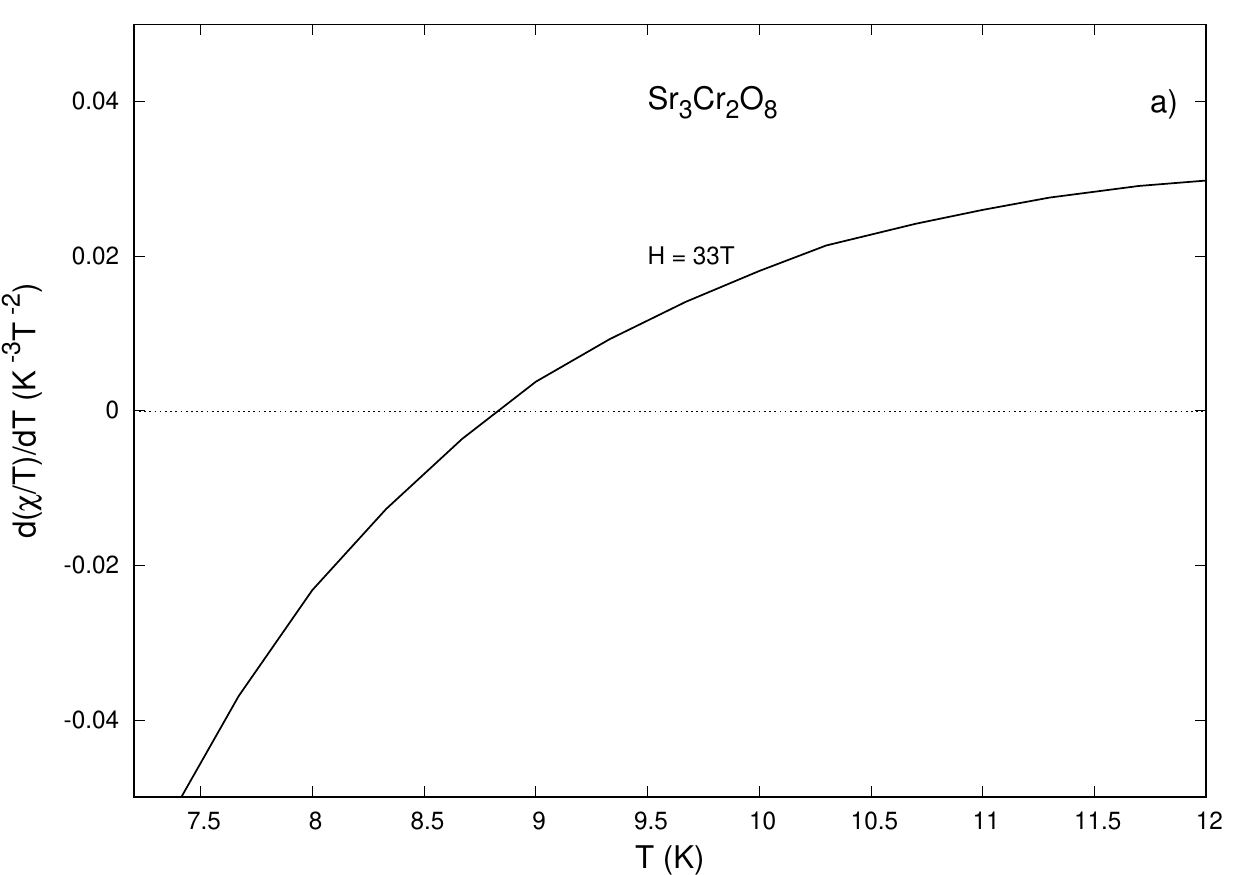} \\ a)}
	\end{minipage}
	\hfill
	\begin{minipage}[H]{0.49\linewidth}
		\center{\includegraphics[width=1.0\linewidth]{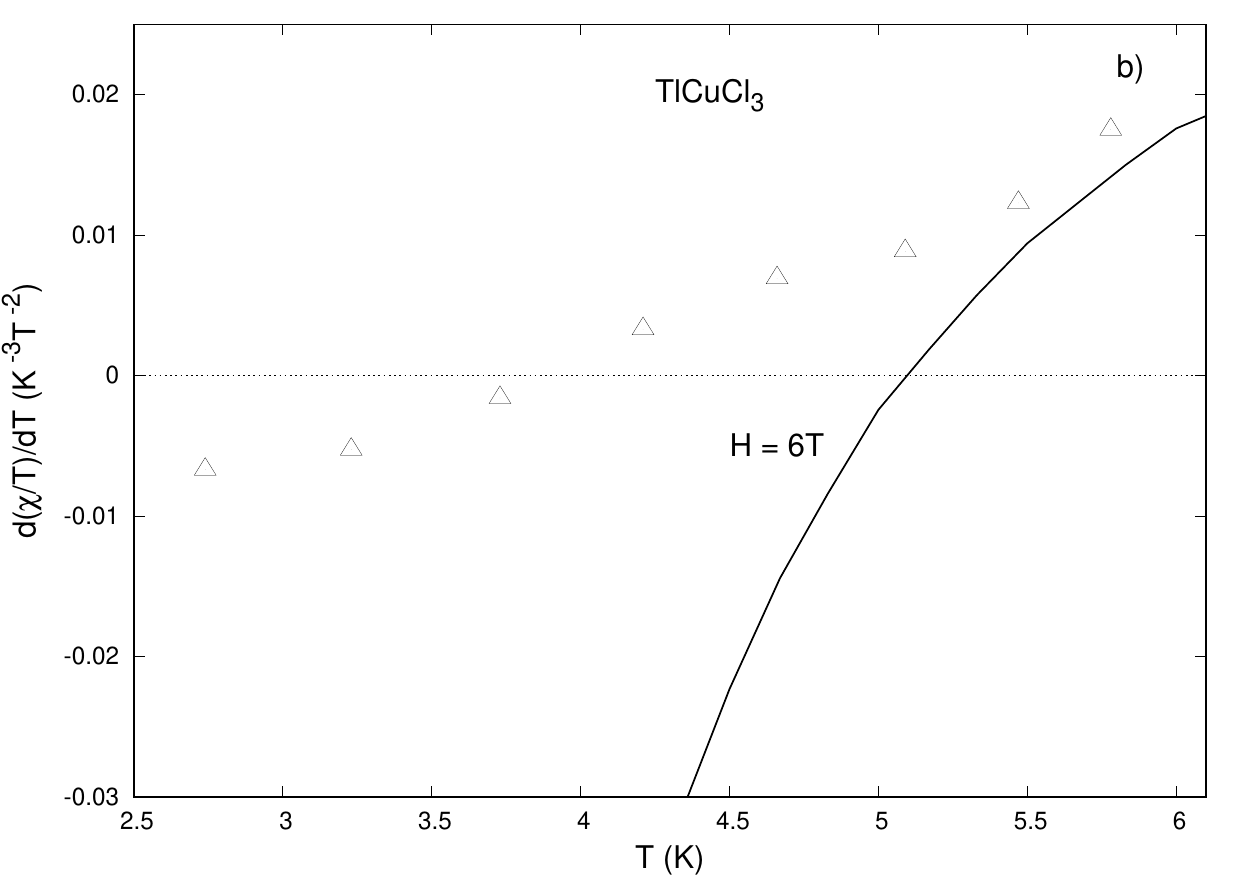} \\ b)}
	\end{minipage}
	\caption{The quantity $d(\chi/T)/dT$ vs temperature  for Sr$_3$Cr$_2$O$_8$ (a) and TlCuCl$_3$ (b). The point where it changes its sign corresponds to the inversion temperature. The triangles in Fig.\,2b correspond to  $d(\chi/T)/dT$
	extracted from the experimental data on $M(T)$ for TlCuCl$_3$ \protect\cite{delamore}.}
	\label{fig2}
\end{figure}

\begin{figure}[h!]
	\begin{minipage}[H]{0.49\linewidth}
		\center{\includegraphics[width=1.0\linewidth]{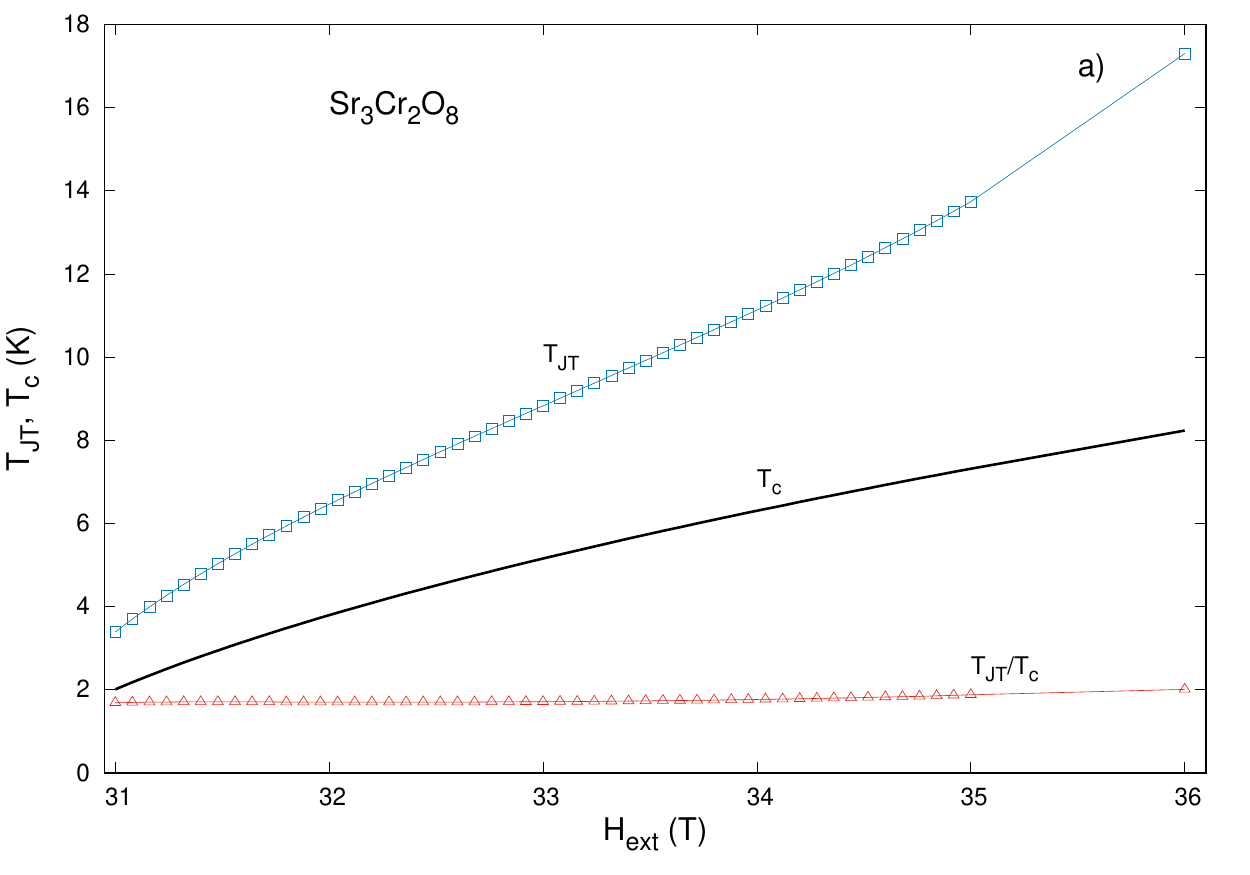} \\ a)}
	\end{minipage}
	\hfill
	\begin{minipage}[H]{0.49\linewidth}
		\center{\includegraphics[width=1.0\linewidth]{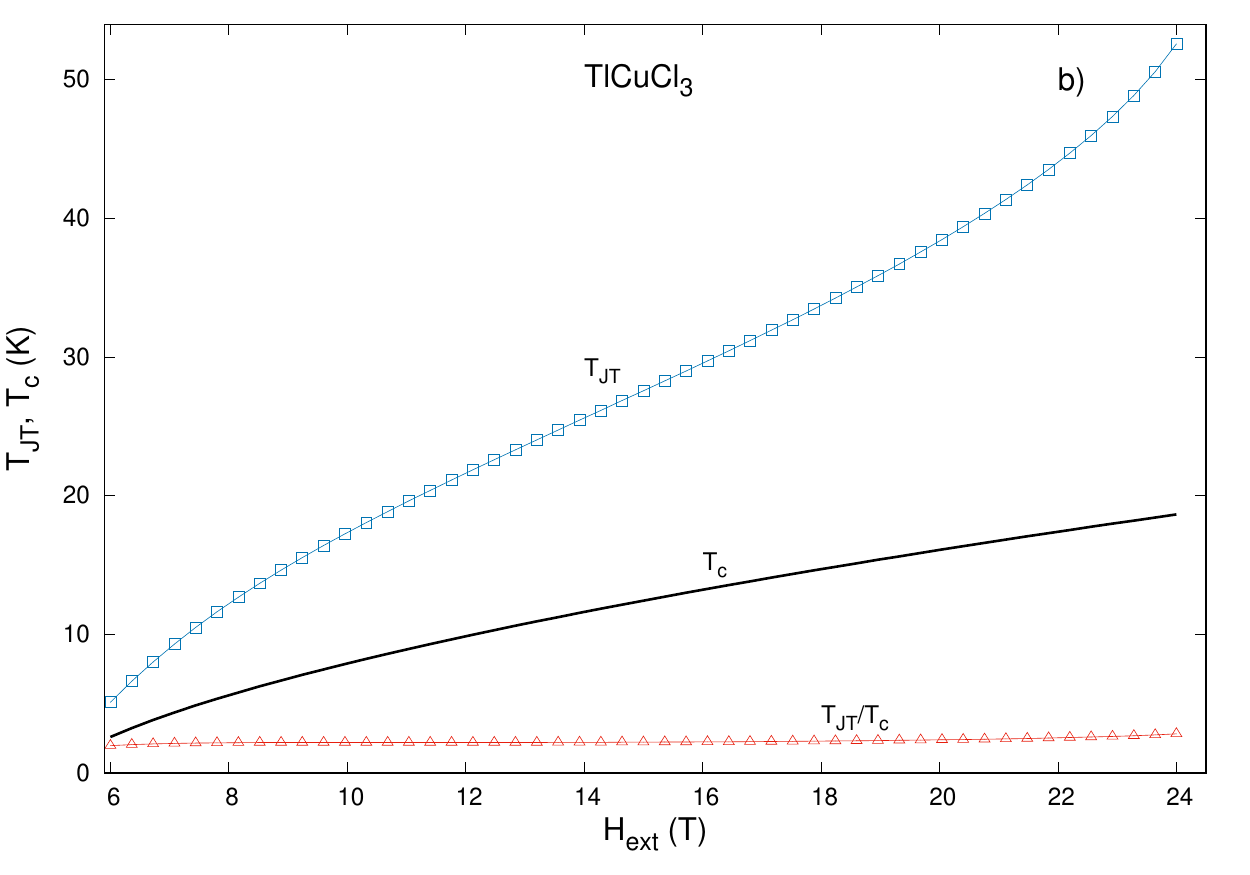} \\ b)}
	\end{minipage}
	\caption
	{
		The magnetic field dependence of the inversion temperature $T_{JT}$  (solid), critical temperature $ T_c$ (dashed) and the ratio $T_{JT}/T_C$ (dotted curves) for Sr$_3$Cr$_2$O$_8$ (a) and TlCuCl$_3$ (b)
	}
	\label{fig3}
\end{figure}

As it was mentioned in the Introduction the Dresden group \cite{wangprl116}
have been performing measurements for  $Sr_3Cr_2O_8$ in the temperature region $T>T_c$. Particularly, they have observed that, in the region of temperatures $8\,\hbox{K}\leq T < 18\,\hbox{K}$  the sound velocity, and hence bulk modulus have an anomaly which disappears at $ T^*\sim 18K$ (c.f. Erratum for \cite{wangprl116}).
Following their interpretation this fact may provide experimental evidence of for the existence of a field induced triplon liquid in the 3D spin - dimerized quantum antiferromagnet  $Sr_3Cr_2O_8$, and the temperature $T^*$ is a maximal temperature of liquefaction. So, proceeding with the analogy of atomic and triplon gases one may come to the conclusion that the inversion temperature $T_{JT}$ under consideration is nothing but the temperature  $T^*$ found in their work. Actually, as it is seen from Fig.\,3(a) the predicted  Joule-Thomson temperature is $T_{JT}^{max}=17.5$\,K (at $H=36$\,T), which in a good agreement with the experimental $ T^* \sim 18$\,K.

In present model the sound velocity $c$ at $T>T_c$ can be evaluated by using \footnote{see Appendix B}
\begin{equation}
mc^2|_{T>T_c}=2U\rho|_{T>T_c}=\frac{B}{\rho},
\label{mc2}
\end{equation}
where $\rho$ is the density of triplons,  and $B=V\left({\partial^2 F}/{\partial V^2}\right)_{T,N}$ is the bulk module. In Figs.4 (a,b) we plotted dimensionless quantity $mc^2(T)/T$ vs temperature. It is seen that it has a minimum exactly at $T=T_{JT}(H)$ for each $H$ in accordance with experimental  predictions of Ref. \cite{wangprl116}.

And for completeness, as to $T_J$, given by equation ${d(\chi \,T)}/{dT}=0$ we failed to find its solution for finite $T$. Therefore, the system of interacting triplons cannot be considered as quasi-ideal at any temperature.
\begin{figure}[h!]
	\begin{minipage}[H]{0.49\linewidth}
		\center{\includegraphics[width=1.0\linewidth]{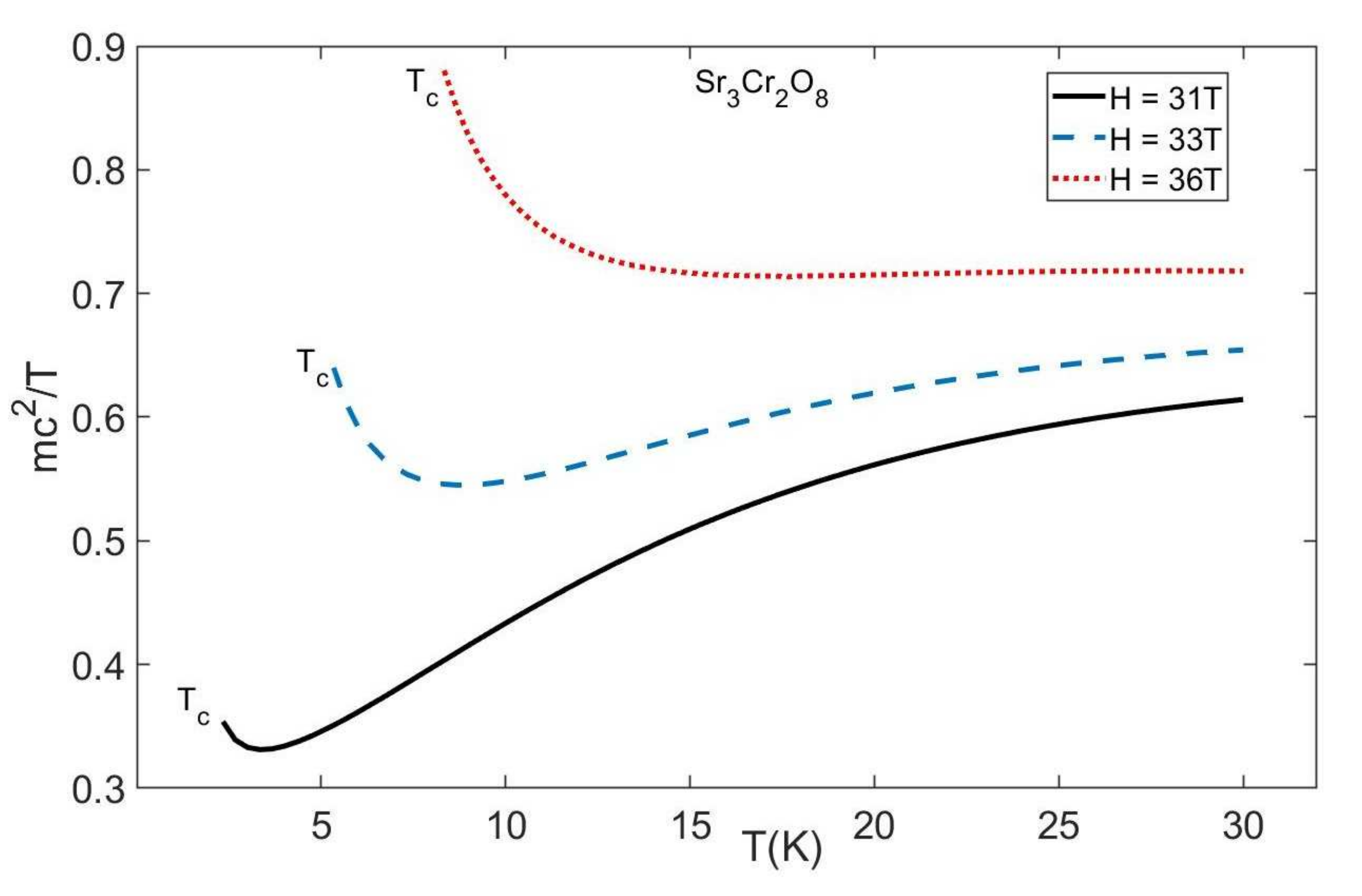} \\ a)}
	\end{minipage}
	\hfill
	\begin{minipage}[H]{0.49\linewidth}
		\center{\includegraphics[width=1.0\linewidth]{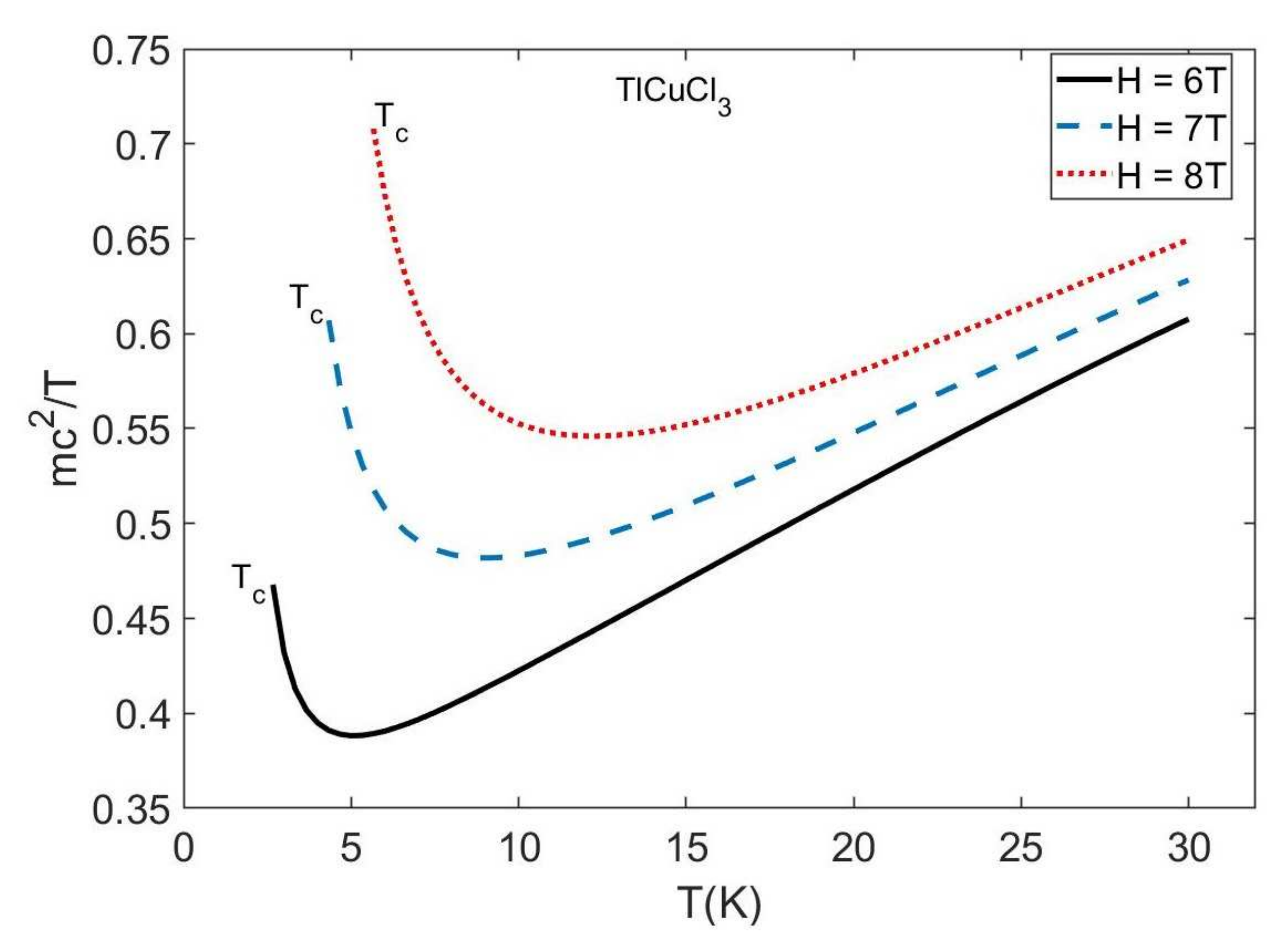} \\ b)}
	\end{minipage}
	\caption
	{
		The  temperature dependence of  $mc^2/T$  in the normal phase $T>T_c$ given by the equation (\ref{mc2}), where c- is the sound velocity. It is seen that
		this quantity, and hence a bulk module  has a minimum near the inversion temperature.
	}
	\label{fig4}
\end{figure}

\section{Conclusion}

We have utilized the BEC analogy to study magnetic thermodynamics of dimerized $s=1/2$ quantum magnets. For this purpose we derived explicit expressions for the characteristic temperatures of dimerized quantum magnets within the Hartree-Fock-Bogoliubov approximation. These equations, as well as experimental
data,  have shown that when the external magnetic field exceeds a critical one , $H>H_c$ the system of triplons has at least two finite characteristic temperatures:  $T_{JT}$ and  $T_c$. The former presents a signature of the liquid state in a temperature region $T\leq T_{JT}$ , while the latter which corresponds to the critical temperature of BEC, shows also the point when in the  triplon spin-liquid a finite superfliud component arises. In this sense, the present work gives an additional argument in order to affirm that the field induced  triplons in 3D spin-dimerized antiferromegnets could be in the liquid state in the range of temperatures $T\leq T_{JT}$, where the Joule Thomson temperature $T_{JT}$ is finite
and of the order of the critical temperature of BEC, $T_{JT}\sim 1.8 T_c$.

Unfortunately,  the present simple approach  cannot describe saturation effects,
since they  are not included in the starting effective Hamiltonian (\ref{ham}) properly. Besides,  for simplicity anisotropic effects, which are essential \cite{sirker,ouraniz1,ourann2} for TlCuCl$_3$  due to Dzyaloshinsky - Moriya (DM) or exchange anisotropy (EA) interactions are neglected. Nevertheless, our predictions on the inversion temperature are in a good qualitative agreement with the existing experimental observations.

\section{Acknowledgements}

We are indebted to Adam Aczel, Zhe Wang ,Vyacheslav Yukalov and   Sergey  Zherlitsyn
for discussions and useful communication. This work is partially supported by  TUBITAK-BIDEB 2221, TUBITAK -ARDEB 1001 programs, Ministry of Innovative Development of the Republic of Uzbekistan and TUBA.

\newpage
\section*{Appendix A }
\def\theequation{A.\arabic{equation}}
\setcounter{equation}{0}
Here we derive equations (\ref{eq62}) and (\ref{eq72}) explicitly. From Eq.s (\ref{eq61}) one may get
\begin{eqnarray}
\left(\frac{\partial E}{\partial M}\right)_T&=&
\left(\frac{\partial (F+TS)}{\partial M}\right)_T\nonumber\\
&=&
\left(\frac{\partial F}{\partial M}\right)_T+
S\left(\frac{\partial T}{\partial M}\right)_T+
T\left(\frac{\partial S}{\partial M}\right)_T\nonumber\\
&=& H+T\left(\frac{\partial S}{\partial M}\right)_T\label{eq63}
\end{eqnarray}
where we used relations (\ref{eq52}) and (\ref{eq61}).
It is clear that
\begin{eqnarray}
\left(\frac{\partial S}{\partial M}\right)_T&=&
-\frac{\partial}{\partial M}
\left(\frac{\partial F}{\partial T}\right)_M\nonumber\\
&=&
-\frac{\partial}{\partial T}
\left(\frac{\partial F}{\partial M}\right)_T=
-\left(\frac{\partial H}{\partial T}\right)_M\label{eq64}
\end{eqnarray}

Now, using (\ref{eq64}) in (\ref{eq63}) one obtains
\begin{eqnarray}
\left(\frac{\partial E}{\partial M}\right)_T \Big|_{T=T_J}&=&H-T_J\left(\frac{\partial H}{\partial T}\right)_M\Big|_{T=T_J}=0\, .\label{eq65}
\end{eqnarray}
This shows that, near $T\sim T_J$ the energy does not dependent on the magnetization.

Now we derive explicit expression for $\kappa_{JT}$ given by Eq. (\ref{eq72}).
Indeed, starting from
\begin{eqnarray}
\kappa_{JT}&=&\left(\frac{\partial T}{\partial H}\right)_W\\
&=&
\frac{\frac{\partial (T,W)}{\partial (H,T)}}{\frac{\partial (H,W)}{\partial (H,T)}}=\frac{\left(\frac{\partial T}{\partial H}\right)_T\left(\frac{\partial W}{\partial T}\right)_H-
	\left(\frac{\partial T}{\partial T}\right)_H
	\left(\frac{\partial W}{\partial H}\right)_T}{\left(\frac{\partial H}{\partial H}\right)_T
	\left(\frac{\partial W}{\partial T}\right)_H-
	\left(\frac{\partial H}{\partial T}\right)_H
	\left(\frac{\partial W}{\partial H}\right)_T} \nonumber\\
&=&-\frac{1}{C_H}\left(\frac{\partial W}{\partial H}\right)_T\, ,\label{eq73}
\end{eqnarray}
and using Eq.\,(\ref{eq61}) it is easy to show that
\begin{eqnarray}
\left(\frac{\partial W}{\partial H}\right)_T&=&T
\left(\frac{\partial S}{\partial H}\right)_T - M\label{eq81}
\end{eqnarray}
and
\begin{eqnarray}
\left(\frac{\partial S}{\partial H}\right)_T&=&-\frac{\partial}{\partial H}
\left(\frac{\partial \Phi}{\partial T}\right)_H =
-\frac{\partial}{\partial T}
\left(\frac{\partial \Phi}{\partial H}\right)_T\nonumber\\
& =&
\left(\frac{\partial M}{\partial T}\right)_H\, .
\label{eq82}
\end{eqnarray}
Inserting (\ref{eq81}) and (\ref{eq82}) into (\ref{eq73}) finally gives $\kappa_{JT}$ in (\ref{eq72}).

\section*{Appendix B}
\def\theequation{B.\arabic{equation}}
\setcounter{equation}{0}

Here we briefly present explicit expressions for  the free energy,
obtained in our earlier work
\cite{ourmce} using a variational perturbative theory
\cite{ourKL,ouryee}. They can be  used for derivation of working fomulas brought in the main text. 
So, in the normal $T>T_c$ and ordered $T\leq T_c$ phases 
the grand thermodynamic potential for triplons is given by
\begin{equation}
\Omega(T>T_c)=-UN^2+T\sum_{k}\ln(1-e^{-\beta \omega_k    })
\label{omegatbig}
\end{equation}
and
\begin{eqnarray}
\Omega(T\leq T_c)=\frac{1}{2}\sum_{k}({E}_k-\varepsilon_{k})+T\sum_{k}\ln(
1-e^{-\beta {E}_k})\nonumber\\
\qquad\qquad+U\rho_1(\rho_1-2N)-\frac{\Delta^2}{2U}
\end{eqnarray}
where
\begin{equation}
 \Delta=\mu+2U(\sigma-\rho_{1}),
 \label{delta}
  \end{equation}
\begin{eqnarray}
  \sigma&=&-\Delta\sum_{k}\frac{W_{k}}{E_{k}},\label{11.6}\\
  \rho_{1}&=&\sum_{k}\left[\frac{W_{k}(\varepsilon_{k}
   +\Delta)}{E_{k}}-\frac{1}{2}\right],
   \label{sigrho1}
    \end{eqnarray}
    with
$W_{k}=\frac{1}{2}\coth\left(\frac{\beta {E}_k}{2}\right)$,
$E_k=\sqrt{\varepsilon_{k}(\varepsilon_{k}+2\Delta)}$.

Now we bring explicit expressions for $\mathcal{E}'_{k,T}=
\left({\partial \mathcal{E}_k}/{\partial T}\right)_H$ and
$\mathcal{E}'_{k,\mu}=
\left({\partial \mathcal{E}_k}/{\partial \mu}\right)_T$
which were used to calculate
$C_H$   and $\Gamma_H$ in the Section III.

 In the normal phase when $\mathcal{E}_{k}=\omega_{k}=\varepsilon_{k}
  -\mu+2U\rho$, the density of particles is given by
  \begin{equation}
   \rho=\sum_{k}f_{B}(\omega_{k})\label{a1}
     \end{equation}
  where $f_{B}(x)=1/(e^{\beta x}-1)$.
  Clearly,
  \begin{equation}
  \frac{d\omega_{k}}{dT}=2U\frac{d\rho}{dT}\label{a2}
   \end{equation}
   which does  not depend on momentum $k$.
Differentiating both sides of the equation (\ref{a1}) with
  respect to $T$ and solving by $dp/dT$, we find
\begin{eqnarray}
 \frac{d\rho}{dT}&=&\frac{\beta S_{1}}{(2S_{2}-1)},\nonumber\\
 \\
  S_{1}&=&-\beta \sum_{k}\omega_{k}f_{B}^{2}(\omega_{k})
   e^{\omega_{k}\beta},\nonumber\\
   S_{2}&=&-U\beta\sum_{k}f_{B}^{2}(\omega_{k})
   e^{\omega_{k}\beta}.\label{a3}
  \end{eqnarray}
Taking the derivative with respect to $\mu$ gives
  \begin{eqnarray}
  \frac{d\omega_{k}}{d\mu}=2U\frac{d\rho}{d\mu}-1,\nonumber\\
   \frac{d\rho}{d\mu}=\frac{S_{2}}{U(2S_{2}-1)}.\label{a4}
    \end{eqnarray}
In the condensed phase, $T \leq T_{c}$, $\mathcal{E}_{k}=
E_k=\sqrt{\varepsilon_{k}(\varepsilon_{k}+2\Delta)}$, and
hence we
  have
  \begin{eqnarray}
  \frac{dE_{k}}{dT}=\frac{\varepsilon_{k}}{E_{k}}
   \Delta_{T}^{\prime},\nonumber\\
   \frac{dE_{k}}{d\mu}=\frac{\varepsilon_{k}}{E_{k}}
   \Delta_{\mu}^{\prime}.\label{a5}
   \end{eqnarray}

   To find, e.g., $\Delta_{T}^{\prime}$ we can differentiate
   both sides of the equation (\ref{delta}) with respect to
   $T$ and solve it for $\Delta_{T}^{\prime}$.

   The results are
   \begin{eqnarray}
    &&  \Delta_{T}^{\prime}=\frac{d\Delta}{dT}=\frac{g S_4}
   {2T(2S_{5}+1)},\nonumber \\
   && \Delta_{\mu}^{\prime}=\frac{d\Delta}{d\mu}=\frac{1}
   {2S_{5}+1},\nonumber \\
   && S_4=\sum_{k}W_{k}^{\prime}(\varepsilon_{k}+2\Delta),\\
&& S_{5}=U\sum_{k}\frac{4W_{k}+E_{k}W_{k}^{\prime}}{4E_{k}},\nonumber
\label{bdelta}
\end{eqnarray}
where
\begin{eqnarray}
W_{k}^{\prime}&=&\beta(1-4W_{k}^{2}).
\end{eqnarray}
As to the  equation (\ref{mc2}), which holds fot $T>T_c $, it can be derived from following equations, proven by Yukalov in Ref. \cite{yukalovtutor}
\begin{equation}
\begin{array}{l}
mc^2=\left(\dsfrac{\partial P} {\partial \rho}\right)\\
\kappa_{T}=\dsfrac{V}{\rho^2}\left(\dsfrac{\partial ^2 F}{\partial \rho ^2}        \right)^{-1}
\end{array}
\end{equation}
where $\kappa_{T}$ is the isothermal compressibility, and $F=\Omega+\mu N$ with $\Omega$ is given by \ref{omegatbig}


\end{document}